%% file: journal_manuscript.tex
\documentclass[3p]{elsarticle}

\usepackage[
    type={CC},
    modifier={by-nc-nd},
    version={3.0},
]{doclicense}
\usepackage[centertags]{amsmath}
\usepackage{latexsym}      
\usepackage{graphicx}      
\usepackage[page]{appendix}
\usepackage{enumitem}
\usepackage[colorlinks=true, linkcolor=black, citecolor=blue, urlcolor=blue, pdfauthor=author]{hyperref}
\usepackage{caption}
\usepackage{tabulary, booktabs, multirow}

 \biboptions{sort&compress}
 
 \journal{Physica A}
\begin{document}
 	\doclicenseThis

 	\begin{frontmatter}
 		
 		\title{Crossover transitions in a bus--car mixed-traffic cellular automata model}
 		
 		\author{Damian N. Dailisan\corref{cor1}}
 		\ead{ddailisan@nip.upd.edu.ph}

 		\author{May T. Lim}
 		\ead{may@nip.upd.edu.ph}
 		\address{National Institute of Physics, University of the Philippines Diliman, 1101 Quezon City, Philippines}
 		
 		\cortext[cor1]{Corresponding author}

 		\begin{abstract}
             \input{abstract.tex} 
 		\end{abstract}
 		\begin{keyword}
 			Computer modeling and simulation \sep Transportation \sep Monte Carlo methods statistical physics and nonlinear dynamics 
 		\end{keyword}
 	\end{frontmatter}

\input{1-introduction.tex}
\input{2-model.tex}
\input{3-results.tex}

\input{4-summary.tex}
\section*{Acknowledgment}
The authors acknowledge the support provided by the Commission on Higher Education-PCARI [IIID-2016-006].

\bibliography{library} 

\end{document}

%% file: abstract.tex
We modify the Nagel-Schreckenberg (NaSch) cellular automata model to study mixed-traffic dynamics. We focus on the interplay between passenger availability and bus-stopping constraints. Buses stop next to occupied cells of a discretized sidewalk model. By parametrizing the spacing distance between designated stops, our simulation covers the range of load-anywhere behavior to that of well-spaced stops. The interplay of passenger arrival rates and bus densities drives crossover transitions from platooning to non-platooned (free-flow and congested) states.
We show that platoons can be dissolved by either decreasing the passenger arrival rate or increasing the bus density. The critical passenger arrival rate at which platoons are dissolved is an exponential function of vehicle density. We also find that at low densities, spacing stops close together induces platooned states, which reduces system speeds and increases waiting times of passengers.

%% file: 1-introduction.tex
\section{Introduction}
\label{sec:intro}
Public transportation is universally acknowledged as a fundamental component in solving traffic congestion.
Together with rail systems, buses form the backbone of medium- to long-haul modes of people transport. 
Since the interaction of buses and passengers introduces complex behavior in transportation systems, more so in cities that do not have designated stops, several models have been proposed to study bus traffic \cite{Oloan1998,Luo2012,Kieu2019,Chowdhury2000a, Nagatani2001}.
In these models, a delay in the arrival time of a bus leads to more passengers waiting for the bus, which then leads to further delays.
Succeeding buses find fewer waiting passengers leading them to catch up to the delayed bus.
Thus, buses form platoons (or bunches).
These models also show a transition from the platooning state to a non-platooned state with increasing bus density.

The tendency of buses to form platoons is problematic for public transport.
In an ideal scenario, an efficient transport system would try to maintain equal time intervals between vehicle arrivals.
However, an equal headway configuration of vehicles is unstable \cite{Helbing1997}, more so with buses \cite{Oloan1998}. Since the instability is inherent to the interaction between public transport vehicles and passengers, approaches to maintain equal headways should consider both traffic and passenger behavior \cite{Gershenson2009}.

Bus route models \cite{Oloan1998,Luo2012,Kieu2019} omit interactions between buses and other vehicle types. 
Yet we expect that these vehicular interactions play a key role in the dynamics of traffic flow --- buses making curbside stops impede traffic flow outright \cite{Nguyen-Phuoc2018}, while small perturbations in vehicle speeds can induce congestion \cite{Bando1995, Sugiyama2008}.
Even a single bus in two-lane mixed traffic alters traffic states and jam transition densities \cite{Nagai2005a}.
As such, it is also critical to decision makers when curbside stops have to be replaced by bus bays to alleviate congestion \cite{Koshy2005}.
Using a modified comfortable driving model, Yuan et al. \cite{YUAN2008} focused on system performance to show a dependence of the system capacity on the number of bus stops.
They also found a gradual transition from platooned to non-platooned states.
However, they did not explore the interplay of passenger arrival rates and vehicle densities extensively nor considered passenger waiting times. 

It is important to understand transitions in traffic models with public transportation in mind.
Slower moving buses and the associated platoon formation can result in the perception that buses cause congestion.
Even a few undisciplined bus drivers can often cause traffic jams when their large vehicles straddle multiple lanes \cite{Dailisan2016, Dailisan2019}.
Such situations, as well as a lack of understanding of these systems, can lead to proposals by policymakers to ban buses to alleviate congestion.
An example of which would be the proposal to ban provincial buses (which account for less than 3\% of traffic volume \cite{MMDA2018}) from Metro Manila's highways to alleviate congestion in one of the city's major thoroughfares \cite{Boquet2013, Rappler2019}.
A Melbourne study found that the removal of such bus services would see up to 30\% of bus users shifting to cars, which defeats the purpose of eliminating buses to alleviate congestion \cite{Nguyen-Phuoc2018}.
Knowledge of the factors that influence transitions can inform decisions, avoiding policies that can cause more harm to the current state of traffic congestion.

This work studies the transitions induced by vehicle density and passenger arrival rate on a Nagel-Schreckenberg model modified to include buses.
We show how the interplay between vehicle density and passenger arrival rates affect the crossover transition from non-platooned to platooning states.
Both single- and multi-lane cases, as well as mixed traffic scenarios, are studied.
Lastly, we look at the effects of the placement of bus stops on traffic flow, as well as passenger waiting times.

%% file: 2-model.tex
\section{Modeling bus--car traffic}

\subsection{Modified Nagel-Schreckenberg model}\label{sec:modified}

\begin{figure}[tb]
	\centering
	\includegraphics[width=5.5in]{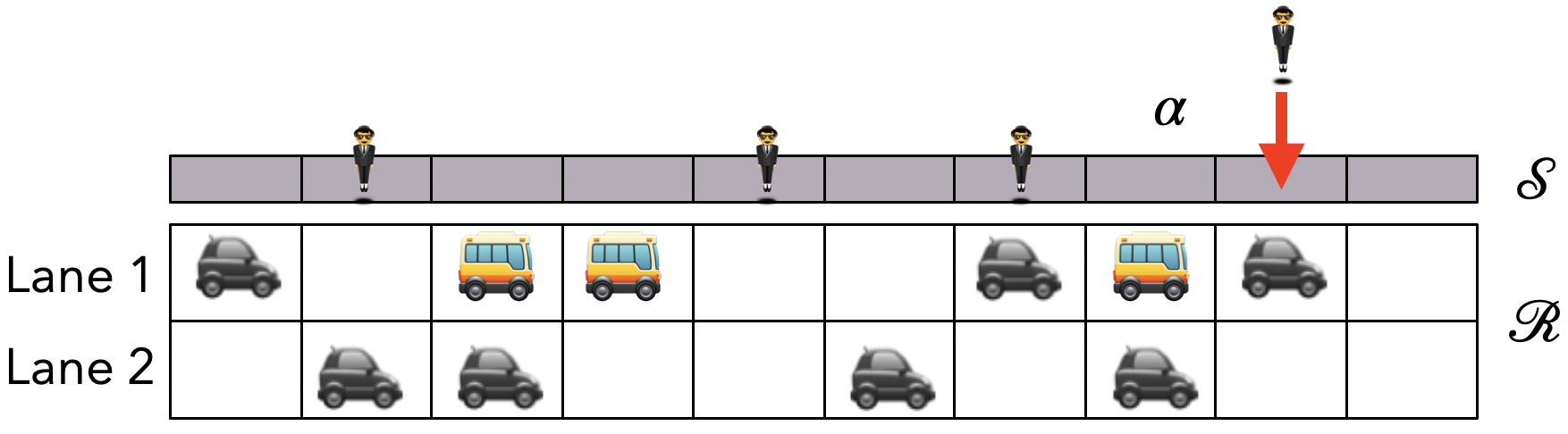} 
	\caption[Diagram of road configuration]{Diagram of road $\mathcal{R}$ and sidewalk $\mathcal{S}$ models.
	Pedestrians can spawn on an empty cell of $\mathcal{S}$ with probability $\alpha$.
	In multi-lane scenarios, buses are limited to the lane adjacent to the sidewalk.}\label{fig:diagram}
\end{figure}

Our model combines elements from the Nagel--Schreckenberg (NaSch) \cite{NaSch1992, Dailisan2016} model and Bus Route Model (BRM) \cite{Oloan1998}.
The two components of this hybrid model are the road and sidewalk (Fig. \ref{fig:diagram}).
The road model $\mathcal{R}$ has $N$ lanes of length $L$ sites, with periodic boundary conditions.
A vehicle can interact with passengers (buses), or ignore passengers (cars).
A sidewalk model $\mathcal{S}$ with one lane of length $L$ sites is updated synchronously with the road model.
States for the $i$th vehicle are lane $l^i$, position $x^i$, and speed $0 \leq v^i \leq v_\mathrm{max}$.
We denote sites occupied by cars as $\mathcal{R}(x,l)=1$ and those occupied by buses as $\mathcal{R}(x,l)=2$.
The road lane adjacent to the sidewalk is denoted as $l=1$.
Passengers occupy the sidewalk if $\mathcal{S}(x)=1$.
Similar to the BRM, we impose that $\mathcal{R}(x,l=1)=2$ and $\mathcal{S}(x)=1$ cannot occur simultaneously, i.e. passengers cannot spawn when a bus occupies the adjacent road cell.

For simplicity, both vehicle types have similar parameters $v_\mathrm{max}$ and $p_\mathrm{slow}$, with the only difference between the two being the additional interaction with pedestrians\footnote{For the sake of brevity and since a passenger waiting for the bus to arrive occupies sidewalk space, we use the terms pedestrian and waiting passenger interchangeably.} for buses.
Additionally, buses have infinite passenger capacity, though we include a short discussion on the effect of finite passenger capacities in Sec. \ref{sec:singlelane}.

Realizations of the model involve assigning vehicle density $\rho$, passenger arrival rate $\alpha$, and bus to vehicle ratio $f_B$.
Vehicle states are updated in random sequential order at each timestep $t$ $(\delta t=1.65\,\mathrm{sec/step})$ following these rules:

\newcounter{count}
\stepcounter{count}

\noindent\textbf{R\thecount: Acceleration}: $v_{t+1}^i = \min\left(v_{t}^i+1, v_\mathrm{max}\right)$ \\
\stepcounter{count}
\noindent\textbf{R\thecount: Bus Loading}: $v_{t+1}^i = 0$ if $\mathcal{S}(x^i_t)=1$;  $\mathcal{S}(x^i_{t})=0$\\
\stepcounter{count}
\noindent\textbf{R\thecount: Lane Change}: $l_i = l_i \pm 1 $ with probability $p_l$ if $v_{t+1}^i>\Delta x^i$\\
\stepcounter{count}
\noindent\textbf{R\thecount: Deceleration}: For cars, $v_{t+1}^i = \min\left(v_{t+1}^i, \frac{\Delta x^i}{\Delta t}\right)$. 

For buses,
\begin{equation*}
   v_{t+1}^i = \begin{cases}
   \min\left(v_{t+1}^i, \frac{\Delta x^i}{\Delta t}\right), & \text{if $\sum\limits_{k=1}^{\lfloor \frac{v_\mathrm{max}}{a_\mathrm{dec}} \rfloor} \left(v_t - k a_\mathrm{dec}\right) > \frac{\Delta x_\mathrm{pass}}{\Delta t}$}\\
   \min\left(v_{t+1}^i, \frac{\Delta x^i}{\Delta t},\max(v_t-a_\mathrm{dec}, a_\mathrm{dec})\right), &\text{if $v_t + a_\mathrm{dec}\geq \frac{x}{\Delta t} \geq a_\mathrm{dec}$}\\
   \min\left(v_{t+1}^i, \frac{\Delta x^i}{\Delta t}, \frac{\Delta x_\mathrm{pass}}{\Delta t}\right), &\text{otherwise.}
   \end{cases}
\end{equation*}

\stepcounter{count}
\noindent\textbf{R\thecount: Random Slowdown}: $v_{t+1}^i = \max\left(v_{t+1}^i-1, 0\right)$ with probability $p_\mathrm{slow}$ \\
\stepcounter{count}
\noindent\textbf{R\thecount: Forward Movement}: $x_{t+1}^i = x_t^i + v_{t+1}^i\Delta t $ \\

\captionsetup[table]{width=\textwidth}
\begin{table}[th!]
	\centering
	\caption[Parameter values and calibration]{Parameters and their values used in the simulation. In this model, each cell is 5.5 meters long and a timestep is 1.65 seconds.}\label{tab:parameters}
	\small
	\begin{tabulary}{\linewidth}{c L L c}
		\toprule
		\textbf{Symbol} & \textbf{Description} & \textbf{Typical Values} & \textbf{Model Value} \\ \midrule
		$L$ 		& Length of the road	& 100\,m (city), 10\,km (highway)& 500, 5120 	\\ 
		$v_{max}$ 	& Maximum speed 		& 20\,-\,120\,km/h (residential, highway)	& 5 (60\,km/h)		\\ 
		$f_B$ 	& Fraction of buses		& 0.03 \cite{MMDA2018}	& 0\,-\,1		\\ 
		$p_{slow}$ 		& Slowdown probability 	& 0.01\,-\,0.5 \cite{NaSch1992, Helbing2001,Combinido2010, Lubeck1998}	& 0.1	\\ 
		$p_{l}$ & Lane change probability & -- & 0 (bus), 1 (car)\\ 
		$a_\mathrm{dec}$ & maximum deceleration & 4\,m/s$^2$							& 2\\ 
		$\rho$ 		& Vehicle density   & 0\,-\,180 vehicles/km		& 0.02\,-\,0.98		\\ 
		$T_{\tau}$ 	& Transient time	& -- & 3000 	\\ 
		$T$ 		& Measurement time	& -- & 3000 	\\ \bottomrule
	\end{tabulary} 
\end{table}

The headway $\Delta x^i$ is the number of empty cells ahead of vehicle $i$ while the passenger headway $\Delta x_\mathrm{pass}$ is the distance from the bus to the closest passenger.
The implementation of \textbf{Deceleration} for buses ensures deceleration rates are limited to $a_\mathrm{dec}$ when picking up passengers \cite{Hou2019, Li2016a}.
The first criterion lets a bus ignore passengers that appear when the bus is passing too fast to safely decelerate.
The second criterion allows a bus to anticipate stopping, and decelerate safely.
For cars, \textbf{Bus Loading} is skipped.
Vehicles may change lanes with probability $p_{l}\in \left\{0,1\right\}$ with equal chances of moving left or right.
Vehicle $i$ attempts to move into adjacent lanes to avoid decelerating, but it may only change lanes if the target lane is empty and it satisfies the safety criteria.
It is safe to change lanes when the trailing vehicle in the target lane can safely decelerate without crashing into vehicle $i$.
The safety criteria can be written as $v^i + (x^i-x^j) \geq v^j - a_\mathrm{dec}$, where $j$ denotes the trailing vehicle.
In the context of overtaking vehicles, \textbf{Random Slowdown} is skipped when a vehicle successfully changes lanes.
Otherwise, the original NaSch rules (\textbf{R1, R4, R5, R6}) are followed.
Through these rules, interactions of vehicles and pedestrians give rise to complex dynamics of traffic flow.

We measure the time averaged flow $q$ of the system, defined as 
${q = \dfrac{1}{T}\dfrac{\sum_{t=1}^{T} \sum_{i} v_t^i \Delta t}{L},}$ where $v_t^i$ is the speed of the $i^{th}$ vehicle at time $t$.
The average speed of vehicles in the system is then $\bar{v} = \dfrac{q}{\rho}.$

We allow for a transient simulation time of $T_{\tau}=3000$ timesteps to remove transient behavior in the data \cite{Combinido2010}.
For all realizations, measurements span $T=3000$ timesteps with fifty trials for each set of parameters $\rho$, $\alpha$, and $f_B$.
A complete list of parameters used in our simulations, with their corresponding calibration factor and real-world values, is given in Table \ref{tab:parameters}.

%% file: 3-results.tex
\section{Results and Discussion}

In Sec. \ref{sec:singlelane}, we focus on the interplay of changing passenger arrival rates, fraction of buses, and transitions for the single lane case.
Section \ref{sec:multilane} extends the model to multiple lanes, with buses being limited to the outermost lane.
Section \ref{sec:periodic_stations} discusses placement of bus stops, and the effect of different bus fractions on the waiting time of passengers.

\input{results/single_lane.tex}

\input{results/multi_lane.tex}

\input{results/periodic_stations.tex}

%% file: results/single_lane.tex
\subsection{Single lane case}
\label{sec:singlelane}

If all vehicles on the road are buses ($f_B=1$), we expect different behaviors at the extreme values of $\alpha$. 
For the case $\alpha \to 0$, passengers do not arrive.
With no passengers to pick up, buses do not slow down, and we recover the original NaSch model.
In the NaSch model, we expect a phase transition to occur at ${\rho_\mathrm{crit}=\frac{1}{v_\mathrm{max}+1}}$, which is the maximum allowable density where vehicles have enough headway to avoid slowing down \cite{NaSch1992, Nagel1993}. For the case of $v_\mathrm{max}=5$,  ${\rho_\mathrm{crit}\approx\frac{1}{6}}$.
Thus, increasing the density beyond $\rho_\mathrm{crit}$ induces a phase transition from free-flowing traffic to congestion. On the other extreme, as $\alpha \to 1$, buses stop after every other timestep.
High values of alpha guarantee that there will be a passenger waiting for a bus at the next cell.
Thus, at low densities, we expect buses in this system to move at an average speed of $\bar{v}=0.5$ sites per timestep.
A transition occurs at $\rho\approx\frac{2}{3}$, which corresponds to the critical density for $v_\mathrm{max}=0.5$.

\begin{figure}[tb]
\centering
\includegraphics{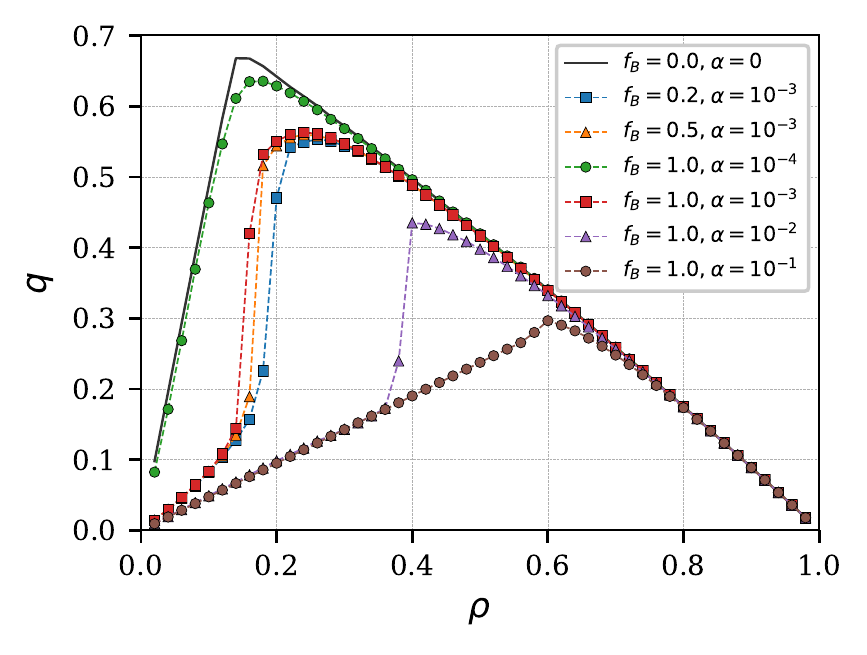}

\caption[Single lane speed vs. density]{
    Fundamental diagram for the single lane case.
    As density increases, jumps in $q$ are observed.
    These jumps occur at different densities, which appear to have a dependence on $\alpha$.
}
\label{fig:fundamental}
\end{figure}

We verify the existence of these two limiting cases (Fig.  \ref{fig:fundamental}).
Two competing trends appear to govern the dynamics of the system:
(1) bus interactions with passengers which tend to drive the average speed of the system to $\bar{v}\to0.5$; and (2) the dynamics of the NaSch model without buses ($\alpha \to 0$ or equivalently, $f_B=0$).
For the values of $f_B=1.0,\,\alpha=[10^{-3}, 10^{-2}, 10^{-1}]$, we observe that for low densities, buses picking up passengers is the dominant behavior.
However, increasing the density of buses results in jumps in throughput in the fundamental diagram.
These jumps indicate that the dominant behavior of the system has shifted towards the dynamics of the fundamental NaSch model (Fig. \ref{fig:fundamental}, $f_B=0$).
We also observe that the density values at which these jumps occur shift to higher densities, as the passenger arrival rates are increased.

This interplay between $\alpha$ and $\rho$ suggests that low values of density allow for passengers to accumulate on different sites of the sidewalk, which forces buses to make more frequent stops.
On the other hand, increasing the density of buses not only accounts for the pick-ups of these passengers, but also prevents the accumulation of passengers in the system.
An interesting consequence is that the average speed of the system actually increases.

\begin{figure}[tb]
\centering
\includegraphics{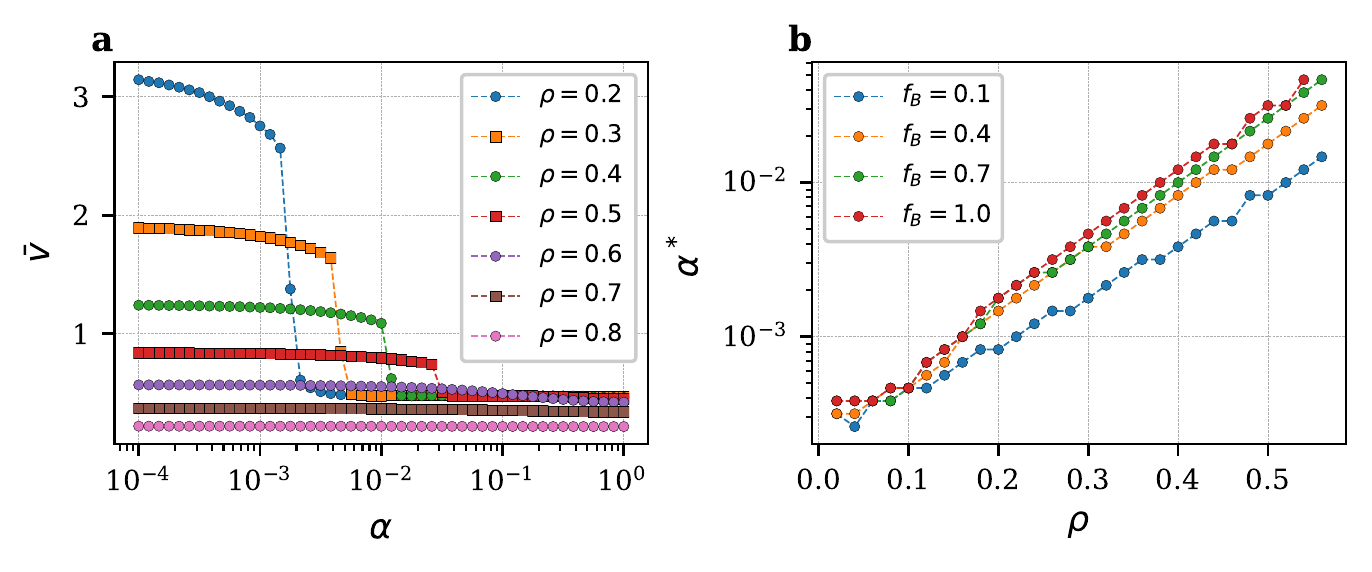}
\caption[Crossover transitions due to arrival rate]{(a) Comparisons of speed as a function of the passenger arrival rate $\alpha$ for different densities ($f_B=1$).
Sudden drops in vehicle speeds as $\alpha$ is increased indicate a crossover from non-platooned to platooned flow.
However, no clear transition is observed for $\rho\geq0.6$.
(b) The crossover transition involves a critical value $\alpha^*$ for different densities and bus fractions. The value of $\alpha^*$ increases monotonically with $\rho$.}
\label{fig:lambda_crossovers}
\end{figure}

\begin{figure}[tb]
\centering
\includegraphics{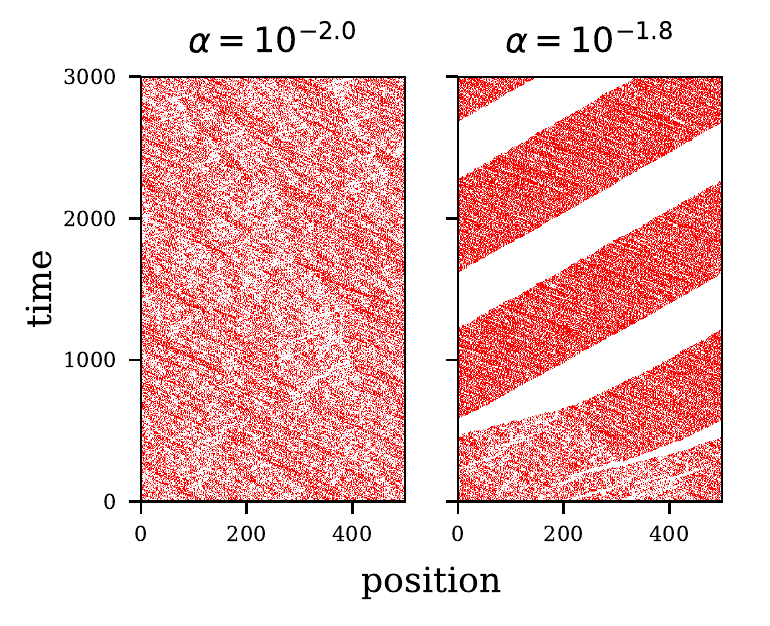}
\caption[Spatio-temporal diagrams near transition point]{Spatio-temporal diagrams of the system of buses ($f_B=1$) near the transition point of $\rho=0.4$, for low passenger arrival rate ($\alpha=10^{-2}$, left) and high passenger arrival rate ($\alpha=10^{-1.8}$, right).
Increasing $\alpha$ induces a transition from non-platooned to platooned flow.
}
\label{fig:std}
\end{figure}

We can look at the interplay between $\rho$ and $\alpha$ in another way, by varying $\alpha$ for fixed values of $\rho$.
We observe that the crossover from non-platooned to platooned states depends on $\alpha$ (Fig. \ref{fig:lambda_crossovers}a).
For a fixed density, once $\alpha$ becomes sufficiently large, passengers accumulate in the system, which results in more frequent stops for buses.
This sets the formation of a platoon of vehicles, while at the same time allowing for more passengers to accumulate ahead of the platoon.

Figure \ref{fig:std} illustrates the crossover from either a free flow or congested phase to a platooned phase for two different values of $\alpha$.
A platoon develops from an initial non-platooned configuration as arrival rates are increased above some critical value $\alpha^*$.
For a system with a density of $\rho=0.4$, we find that
$\alpha^*\approx10^{-1.91}$ (Fig \ref{fig:lambda_crossovers}a).
In the case of $\alpha=10^{-2}$, platoons do not form in the system at all.
However, when we look at the case of $\alpha=10^{-1.8}$, a platoon develops after 600 timesteps, despite having the same initial configuration for the case $\alpha=10^{-2}$.
As passengers accumulate in the gaps between buses, the decrease in the speed of buses creates larger gaps.
This feedback loop drives the formation of platoons and also results in longer passenger waiting times.

For the case of mixed traffic (Fig. \ref{fig:fundamental}, $\alpha=10^{-3}$), 
the transition from a platooned to a non-platooned phase occurs at lower densities with increasing $f_B$.
This observation is still consistent with the notion of the transition from platooned to a non-platooned phase due to increasing density of buses.
As the effective bus density scales with $f_B$, higher densities are needed to compensate for lower $f_B$.
Thus, reducing $f_B$ shifts the crossover transition to higher densities.

We can make two observations of our single lane model.
First, we see that there exists a $\rho^*(\alpha)$ responsible for a crossover behavior from a platooned phase to either a free-flow or congested phase.
When $\rho<\rho^*$, a low $f_B$ gives more time for passengers to accumulate in the gaps between buses.
At the same time, buses will tend to clump together, creating larger gaps.
Both processes aid in the formation of platoons.
When $\rho>\rho^*$, the gaps between buses are sufficiently small to substantially reduce passenger accumulation that cause a cascade of slowdowns.
Thus, smaller or no platoons are formed, and we obtain a non-platooned phase.

Secondly, the crossover from a platooned to a non-platooned phase is also determined by $\alpha^*(\rho)$.
For arrival rates $\alpha<\alpha^*$, the system would be found in the non-platooned phase, while for $\alpha>\alpha^*$ the system will exhibit platooning.
The dependence on $\rho$ is only up to $\rho\approx\frac{2}{3}$, beyond which the density-dependent dynamics of the NaSch dominates the system.
Figure \ref{fig:lambda_crossovers}b highlights the interplay between $\alpha^*$ and $\rho$.
The crossover transition value appears to scale as $\alpha^*\sim\exp(\rho+c)$.
We observe the scaling only until $\rho=0.56$, beyond which the platooning and congestion effects are difficult to distinguish.
Unlike the phase transition from free flow to congestion which only occurs at low densities, this crossover can occur at densities above $\rho_\mathrm{crit}$.
While the phase transition from free flow to congestion is undoubtedly an important aspect in the study of transport and vehicular traffic, the densities involved are typically low densities.
Since the problem of congestion in city traffic involves densities above $\rho_\mathrm{crit}$, this particular phase transition is rather insignificant when it comes to policy intervention since controlling vehicle volume is difficult.
However, the crossover transition of passenger--bus interactions is present for a larger range of densities.
The key to managing traffic flow at densities above $\rho_\mathrm{crit}$ must then lie in avoiding the formation of platoons.

\begin{figure}[tb]
\centering
\includegraphics{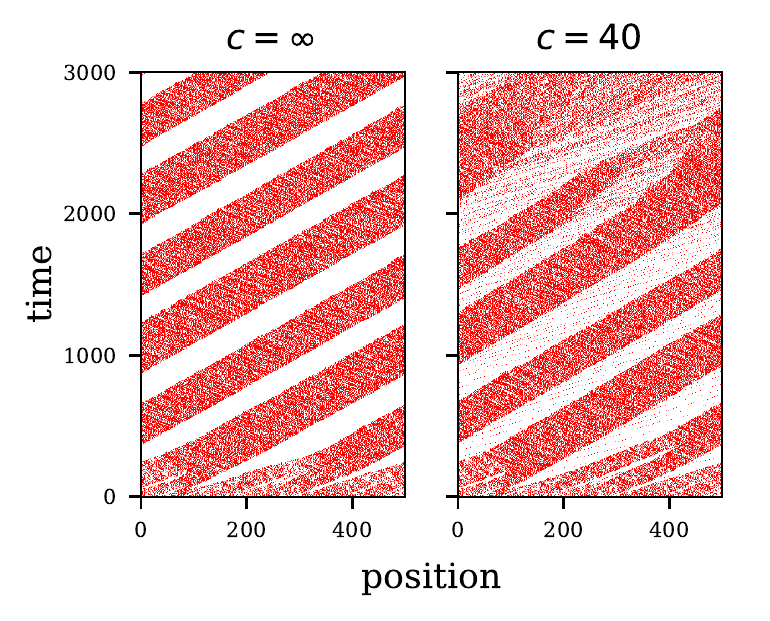}
\caption[Spatio-temporal diagrams of infinite vs. finite bus capacity]{Spatio-temporal diagrams of the system of buses ($f_B=1,\alpha=10^{-1.6},\rho=0.4$) with different passenger capacities $c$.
The system with finite capacity ($c=40$) platoons like our simplified model ($c=\infty$) for some time $T_\mathrm{cap}\approx3000$ until the platoon dissolves as most buses get filled.
Small dots breaking away from the platoon are buses have just reached their capacity.
}
\label{fig:std_capacity}
\end{figure}

The behavior of buses in our simplified model has underlying assumptions that are absent in real traffic.
Buses in our model have infinite capacities, do not wait for passengers, and our model does not have passenger drop-offs.
What happens when we relax these assumptions?

\emph{From infinite to finite bus capacities.}
With each bus having a  passenger capacity $c$, the ``leading bus" gets filled up and eventually stops picking up passengers.
Said bus breaks away from the platoon that it leads, leaving the next bus as the new ``leading bus".
The notion of a ``leading bus" in this case is not specific to a single bus, as any unfilled bus can be the ``leading bus" and cause a platoon. A spatio-temporal diagram of our model with different capacities shows this phenomenon (Fig. \ref{fig:std_capacity}). 

\begin{figure}[tbh]
\centering
\includegraphics[width=\textwidth]{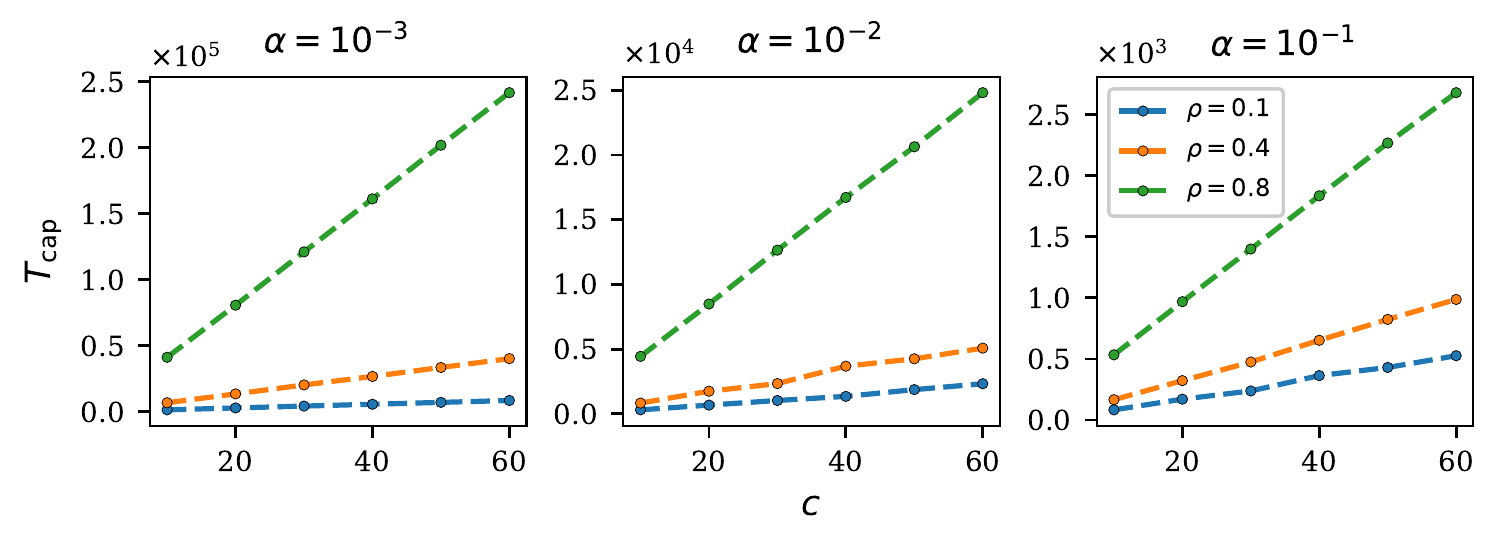}
\caption[]{The time it buses to reach capacity $T_\mathrm{cap}$ as a function of the passenger capacity $c$ for different combinations of $\alpha,\rho$.
$T_\mathrm{cap}$ scales linearly with the passenger capacity for all combinations of density and passenger arrival rate.
}
\label{fig:time_to_fill}
\end{figure}

If the passenger volume eventually exceeds aggregate bus capacity, all buses will be full after some time $T_\mathrm{cap}$.
In the case of Fig. \ref{fig:std_capacity}, $T_\mathrm{cap}\approx3000$.
At that moment, the platoon dissolves, and we recover the dynamics of the NaSch model.
While this speeds up the movement of vehicles and passengers on board the buses, this also means that those who have yet to board will have to wait indefinitely for a bus.
We find that $T_\mathrm{cap}$ scales linearly with the passenger capacity $c$, for all densities and passenger arrival rates (Fig. \ref{fig:time_to_fill}).

\begin{figure}[tb]
\centering
\includegraphics{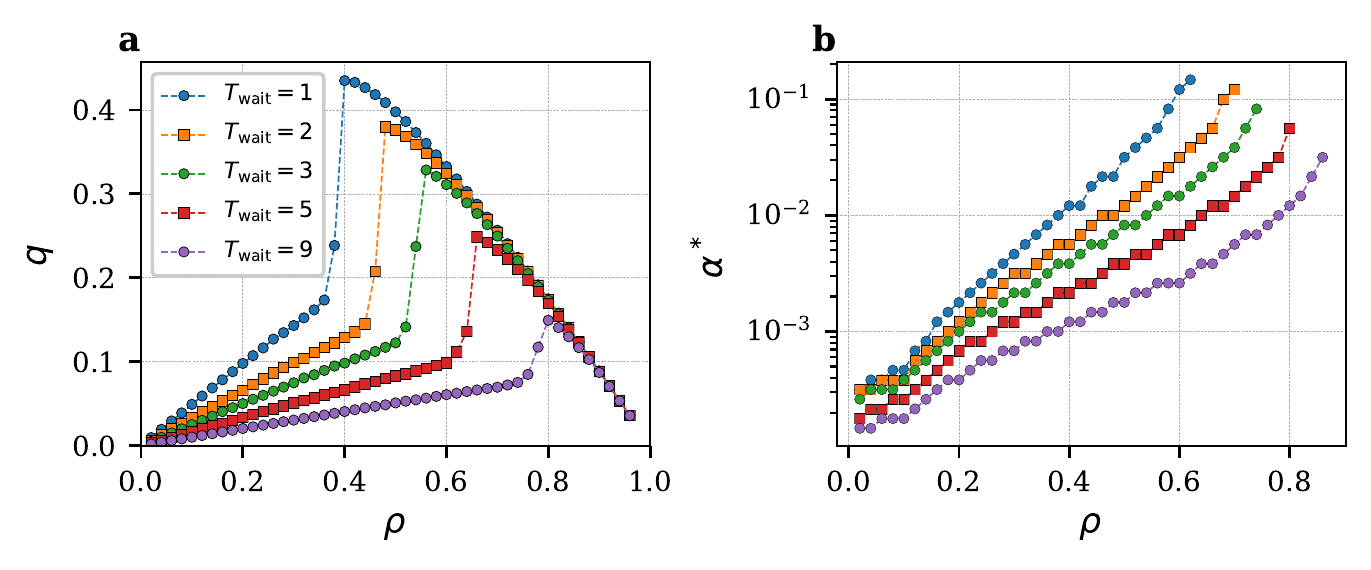}
\caption[]{(a) Fundamental diagram for different bus waiting times $T_\mathrm{wait}$.
Waiting for passengers for a longer duration reduces the average speed of buses in the system, and shifts the crossover transitions to higher densities.
(b) The critical value $\alpha^*$ vs. densities for different bus waiting times. Increasing $T_\mathrm{wait}$ changes the slope of $\log\alpha^*$ vs. $\rho$.
}
\label{fig:bus_waiting}
\end{figure}

\emph{From no-waiting to waiting for passengers.}
Allowing buses in our model to stop for a duration $T_\mathrm{wait}$ is analogous to waiting for passengers or making scheduled stops.
In effect, this slows down the movement of the platoon (Fig. \ref{fig:bus_waiting}a).
In the limit that $\alpha$ is sufficiently large, we can calculate the average speed of the platoon as $\bar{v}=\frac{1}{T_\mathrm{wait}+1}$.
In the case of our simplified model, $T_\mathrm{wait}=1$ and we have $\bar{v}=0.5$.

Aside from slowing down the platoon, increasing the stop duration of buses shifts the crossover transition to higher densities.
We find that for the same densities, having shorter waiting times cause their transitions to occur at higher values of $\alpha$ (Fig. \ref{fig:bus_waiting}b).
This ties into the mechanism of platoon formation, as when buses wait longer at stops, the average speed of the platoon decreases.
For  the same pedestrian spawning rate $\alpha$, this decrease in platoon speeds results in more pedestrians accumulating ahead of the platoon.

\emph{Adding drop-offs.} 
Our description of the model does not explicitly let passengers alight.
While the term \textit{picking up passengers} is useful in creating a mental picture of what goes on in the model, a better interpretation is that these are stopping events as a result of passengers boarding and alighting the bus.
Instead of explicitly modeling multiple passengers boarding (and alighting) buses at the sidewalk, buses can be filled (and emptied) by multiple passengers at a single stop.
This works in the regime where the net flux of passengers do not fill up buses.

When buses are close to capacity, explicitly modeling drop offs influences the dynamics of buses.
In Fig. \ref{fig:std_capacity}, we show that as buses reach their capacities, they break away from the platoon, and can end up at the tail of a different platoon.
If no passengers alight, the bus catches up with the tail of the next platoon.
On the other hand, if a passenger alights within the route, the bus once again becomes a potential platoon starter.
In this way, a bus can alternate between a platoon starter, and a platoon trailer, depending on the distribution of drop-off points of passengers.
If passenger drop-offs are spread uniformly throughout the route, we expect to see several smaller platoons instead of a single large platoon.
Note that some commute routes would have a concentration of pickups at the start of the route, and drop-offs at the end of the route.
This situation would resemble the dynamics of buses at capacity, with faster transit speeds as stops mid route are minimized. 

%% file: results/multi_lane.tex
\subsection{Multi-lane traffic}
\label{sec:multilane}

We now extend the single-lane model to two lanes.
In this model, we restrict buses to drive along the second lane, while cars can move freely on any lane.
As buses stop to pick up passengers, they block the movement of vehicles on the bus lane.
Lane changing allows cars to overtake stopped or slow-moving buses, though the maneuver can cause congestion on the adjacent lane.
Buses cause bottlenecks, which greatly reduce the flow of vehicles in the system.
These bottlenecks are similar to work zone scenarios \cite{Hou2019a}, with the key difference being that the flow reduction is temporary and that the bottleneck is non-stationary.

\begin{figure}[tb]
\centering
\includegraphics[height=1.6in]{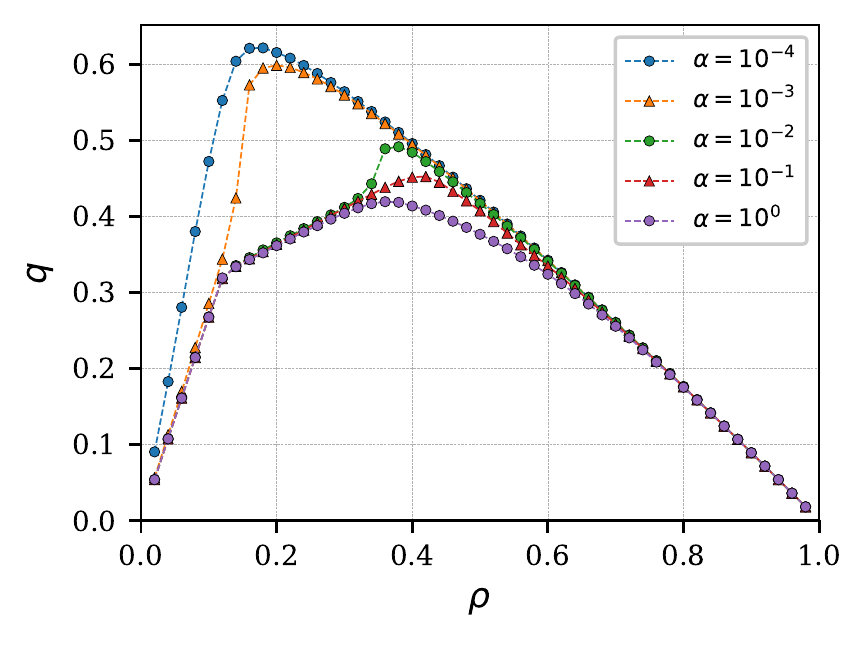}
\includegraphics[height=1.6in]{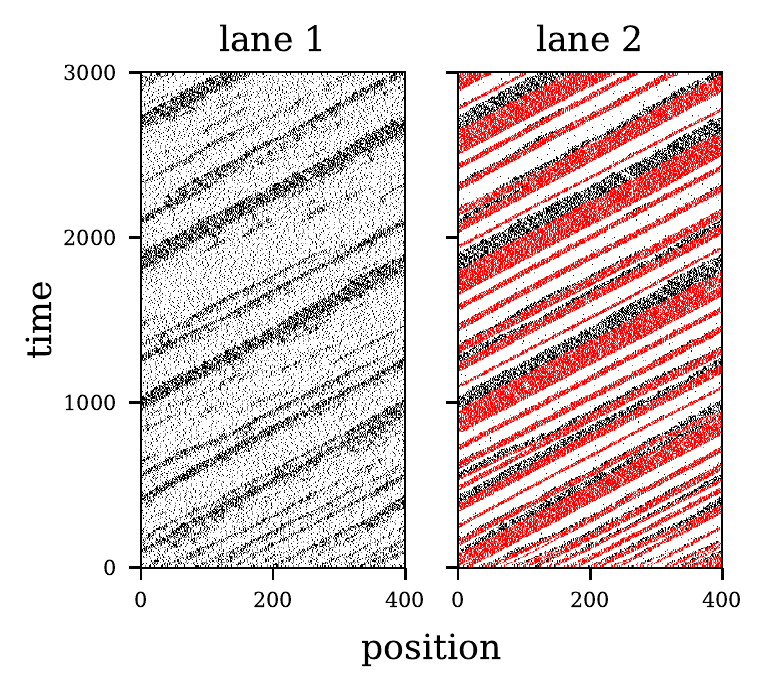}
\caption[Multi-lane fundamental diagram]{
    (Left) Fundamental diagram of multi-lane traffic.
    Jumps in $q$ are not as obvious as in the single lane case, but are still observed for $\alpha=\left[10^{-3},10^{-2}\right]$.
    (Right) Spatio-temporal diagram for $\rho=0.3, \alpha=10^{-1}$, with cars (black) and buses (red).
    Coupling of both lanes occurs at the trailing end of platoons due to cars changing lanes.
}
\label{fig:fd_multi_alpha}
\end{figure}

In the case of multiple lanes, we see three regions in the fundamental diagram for the case of $\alpha\geq10^{-1}$ (Fig. \ref{fig:fd_multi_alpha}).
The first two regions are from $0<\rho<0.12$ and $0.12<\rho<0.66$.
In the single lane case, when $\alpha$ is sufficiently high, a bus moves slowly as it picks up passengers, which obstructs all other vehicles behind it.
In the multiple lane case, cars can overtake the slow-moving buses and prevent the platoon formation ($0<\rho<0.12$).

For the second region ($0.12<\rho<0.66$), traffic flow in the two lanes gets coupled due to cars changing lanes.
As platoons form behind slow-moving buses, they impede the movement of cars in the second lane.
When these cars switch lanes, they also slow down vehicles in the first lane (see spatio-temporal diagram, Fig. \ref{fig:fd_multi_alpha}).
On the other hand, we also see that these cars can utilize gaps between platoons.
Platoons in this multi-lane model do not appear as a single clump of closely spaced vehicles followed by a large gap of space.
Instead, many smaller platoons form, with overtaking cars utilizing the gaps between platoons.
However, this region does not exist for the case of $\alpha\leq10^{-3}$, as the arrival rates of passengers is not high enough to cause the platoon formation.
Thus, in the absence of platoons, the speeds of buses and cars are similar, and we recover dynamics similar to the NaSch model.

The remaining region ($\rho>0.66$) marks the dissolution of the platoons in the bus lane.
In this region, the density has increased enough such that both lanes move at similar speeds.
As a result, the effect of bus pickups is indistinguishable from the congested dynamics of the NaSch model.
We also observe this phase in the case of a crossover transition occurring when $\rho<0.66$ (such is the case when $\alpha=[10^{-2}, 10^{-3}]$).

In this multi-lane scenario, stop durations increase when buses are allowed to wait for passengers.
This slow-down has a two-fold effect: it reduces overall platoon movement; and it impedes the adjacent lane as more cars move out of the bus lane.

%% file: results/periodic_stations.tex
\subsection{Periodic stops}
\label{sec:periodic_stations}

In the previous sections, passengers may board and alight on any segment of the road.
Such a scenario is common in cities like Manila and Jakarta, whereas having designated bus stops is a more common occurrence worldwide.
As such, it is natural to take accessibility into account when designing stop locations.
Closer spacing between stops can improve accessibility for commuters. But when spaced too closely, stops slow down transit travel speeds \cite{Murray2003}.

The decentralization of bus franchise operators in cities like Manila makes it challenging to plan for different aspects of a transit system. In routes that do not have sufficient demand, buses will be infrequent.
Uncertainty in arrival times of buses in such cases can lead to either long waiting times, as well as a reduction of demand as commuters search for a more reliable alternative.
Implementing bus timetables can help manage passenger expectations and establish the reliability of the transit system.

In comparing realizations with different distances between designated stops $d$, we assumed the same total number of passengers. Thus, we fixed the value of the pedestrian arrival rate for the entire system $A = \alpha \left( \frac{L}{d}\right)$ per timestep.
With fewer stops (larger $d$), the arrival of passengers at each stop, $\alpha$, rises commensurately.
We ran simulations for $d\in[1,4,16,64,256]$, and fixed $A=\alpha_\textrm{max}\frac{L}{d_\textrm{max}}$, such that when $d=256$, $\alpha=0.1$.
A maximum value of $d=256$ is specifically chosen for our cell size of 5.5\,m, as this corresponds to a 1.408\,km distance between stops.
We choose this based on the assumption that at most, a pedestrian would be forced to walk 704\,m for a destination in between two stops.
To account for our choice of $d_\mathrm{max}$, we use $L=5120$ and set $A=2$ for simulations in this section.
For comparison, we will look at the cases ranging from all buses ($f_B=1$) to a low bus volume, Philippines-inspired case ($f_B=0.03$) \cite{MMDA2018}.

\begin{figure}[tb]
\centering
\includegraphics[width=\textwidth]{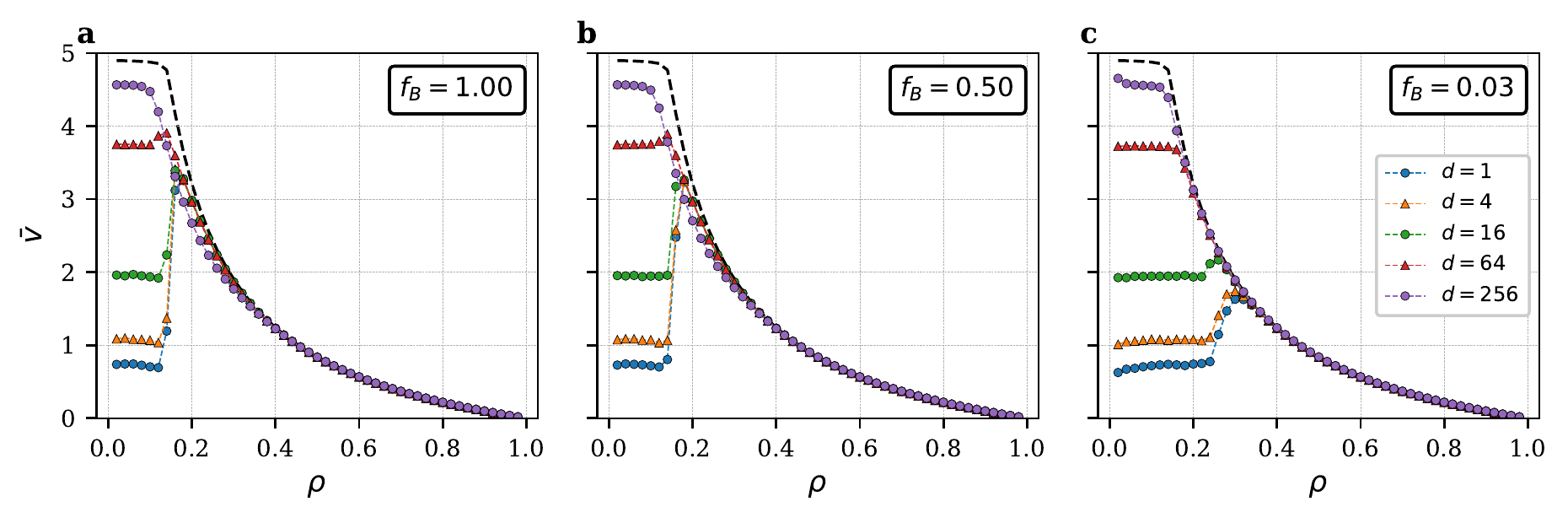}
\caption[Varied station separation distances, speed vs. density]{
Speed vs. density plots for a single-lane model with varying station separation distances $d\in[1,4,16,64,256]$.
Black dashed lines are for the baseline NaSch model with no buses.
Spacing stops further apart increases speeds of vehicles on the system by mitigating platoon formation.
}
\label{fig:fd_stations}
\end{figure}

First, we look at the single-lane case of an all-bus system.
Increasing the distances between stops improves the overall speed of the system for $\rho<0.12$ (Fig. \ref{fig:fd_stations}a and \ref{fig:fd_stations}b).
The inflection point at $\rho\approx 0.12$ corresponds to the crossover transition from platooning to a non-platooned phase.
Configurations with $d\leq64$ exhibit an inflection at $\rho\approx 0.12$, a crossover transition which is absent in farther-spaced stops ($d=256$).
We observe similar behavior in the $f_B=0.03$ case (Fig. \ref{fig:fd_stations}c), where system speeds increase with station separation distances, but with two key differences from the all-bus ($f_B=1$) case.
Only $d=1$ and $d=4$ have crossover transitions that occur at higher densities, and the case of $d=256$ does not have slower system speeds than $d\leq64$ for all densities.

We attribute the slowdown of vehicles in the case of $d=256, f_B=1$ to the high arrival rates of passengers, which form slow-moving jams near the stops where buses have a high probability of stopping.
Due to our model design, buses do not wait for passengers at stops.
Since the distance (and time) headways between succeeding buses are short, the high arrival rates of passengers would make succeeding buses stop at the station.
If a bus can wait for multiple passengers, buses behind it would not have to stop often, which would improve the overall system speeds.

In general, we expect the average number of buses that stop to board passengers should be the same for all realizations of $d$.
Allocating larger separation distances between stops reduces the frequency of stops made by buses.
This allows buses to maintain their top speeds for longer, and results in a higher system throughput.

\begin{figure}[tb]
\centering
\includegraphics{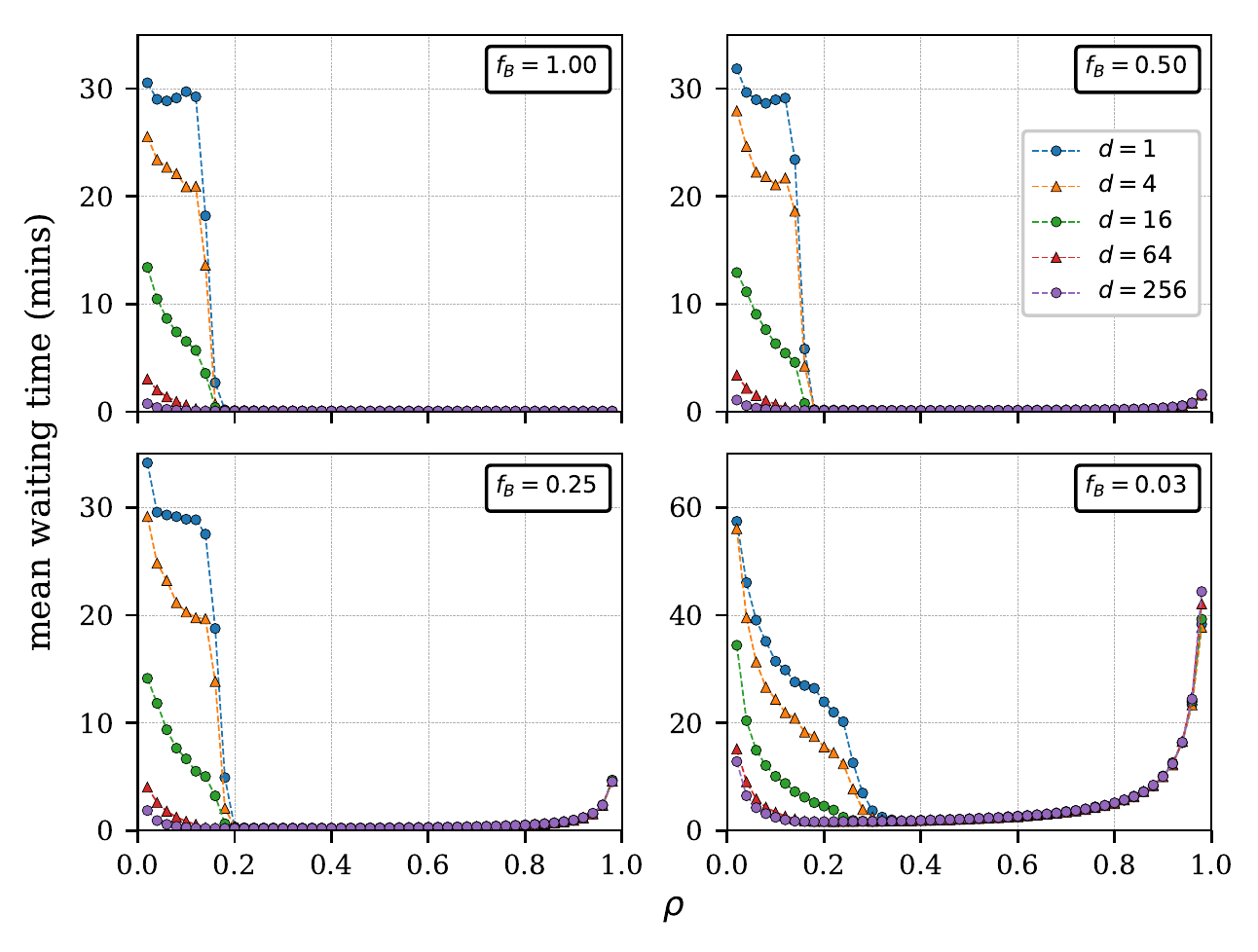}
\caption[Mean waiting times of passengers]{
Mean waiting times at various densities for different station separation distances $d$.
Station distances affect waiting times at middle to low density values ($\rho<0.32$).
Longest waiting times can be found at low densities (all $f_B$), and at high densities for mixed traffic ($f_B<1$).
}
\label{fig:waiting_time}
\end{figure}

System speeds, however, only paint half of the picture.
Travel begins when the pedestrian leaves his origin, not at the moment the pedestrian boards a vehicle.
Thus, it is necessary to include the time spent waiting for public transport.
We find that the spacing of stops matters most in medium to low density regimes ($\rho<0.2$ for $f\in[0.25, 0.5,1]$, $\rho<0.32$ for $f_B=0.03$, Fig. \ref{fig:waiting_time}).
Having stops close to each other increases waiting times of passengers, more so at low densities.
Because of platooning, bus delays lead to more waiting passengers, which in turn causes further delays.
Thus, waiting times can reach more than 30 minutes when buses are allowed to stop anywhere ($d\in[1,4]$).
Increasing stops separation therefore results to faster system speeds and shorter waiting times.

For low densities, passenger waiting times increase with decreasing system density.
Infrequent bus arrivals force passengers to wait longer.
Spacing stops close together exacerbate platooning effects --- even spacing of bus headways are not maintained, and stopping occurs more frequently.
This also leads to inefficiencies, as buses may arrive just after the preceding bus finished boarding.
In situations where public transport is deficient by way of long waiting times, having a bus timetable would help.
A communicated schedule allows passengers to adjust their arrival at the station accordingly, effectively lowering expected waiting times.

For higher densities ($\rho>0.16$), waiting times become constant for the all-bus case (Fig. \ref{fig:waiting_time}, $f_B=1$).
Furthermore, the wait is just one timestep: the pedestrian gets to ride the bus one timestep after getting to the sidewalk.
While this looks good from the point of view of passengers, it is quite inefficient since some buses  fail to pick up passengers.
We set the net pedestrian arrival rate $A=2$, but the number of buses in the system for $\rho>0.18$ is more than 921.
Even if our model does not take into account bus passenger capacities, we can already see that reducing the number of buses may improve system speeds without loss of service reliability.

In mixed traffic settings, waiting times do not remain constant with increasing vehicle density.
At sufficiently high road densities, waiting times also increase.
Increased waiting times at higher road densities are an indication of insufficient supply of buses.
The lack of buses, coupled with the slow speeds of vehicles on the road, lead to prolonged waiting times as congestion gets worse (Fig. \ref{fig:waiting_time}, $f_B=0.03$).

Allowing passengers to board and alight vehicles at any point is an attractive idea.
The convenience of being able to go directly from your origin to your destination is the value proposition of owning personal vehicles or using taxis, ride-sharing services, and forms of para-transit such as tricycles, pedicabs, and \textit{habal-habal}, which are prevalent in the Philippines.
This mentality bleeds into other forms of public transportation such as buses, Jeepneys, and AUVs, which do not have designated stops.
Our work supports the notion of planning out stops that are spaced further apart, but we must also take into account the added time and effort to the commute of a passenger whose stops would lie in between two stops.

The growing popularity of ride-sharing services also causes problems.
Although these have infrequent stops, their point-to-point nature is similar to a system of buses that do not have well spaced stops.
Entire fleets of these vehicles can end up clogging roads and side-streets while they wait for their next booking.
If left unchecked, the sheer number of ride-sharing vehicles waiting for passengers can end up negating any benefits from a well designed public transport system.

%% file: 4-summary.tex
\section{Conclusion}

Platooned flow is characterized by bunched slow-moving buses that leave gaps on the road. Platoons disappear either by having fewer riders or more buses, and results to an equal headway configuration.
With multiple lanes, platooned dynamics spill over to the non-bus lane, and persists until the platoon is dissolved.

Platoon formation is driven by interactions at stops. Spacing out stops speeds up traffic by minimizing platooning and queuing-induced congestion. However, spacing them too far would make walking to a midpoint destination impractical. Load-anywhere behavior, while convenient for an individual, comes at the cost of slower traffic flow and longer waiting times. 
To mitigate the formation of platoons, sticking to a bus schedule helps. Cities may also incentivize staggered work schedules while ensuring adequate bus supply.

Though our model did not take into account multiple passenger arrivals on sidewalk cells, finite bus capacities, and alighting scenarios; we can estimate their effects.
If we account for multiple passenger arrivals, the waiting times belong to the first passenger in the queue, and thus a lower limit.
Adding finite capacities to buses dissolves platoons when buses are full, but the resulting increase in vehicle speeds comes with longer passenger waiting times.
Allowing buses to wait for passengers at stops, rather than the other way around, slows down the overall movement of platoons resulting to congestion.

Microscopic models, like ours, would benefit from the increasing use of real-time sensors that drive smart city initiatives worldwide. With rapid parameter calibration, we can realize the idea of immediately actionable predictions.